\documentclass{article}
\usepackage[numbers]{natbib}
\usepackage[preprint]{neurips_2025}

\usepackage{hyperref}
\usepackage{tcolorbox}
\usepackage{graphicx}
\usepackage{tabu}                 
\usepackage{booktabs}                 
\usepackage{lipsum}                    
\usepackage{mwe}                        
\usepackage{fontawesome5}
\usepackage{tikz}
\usepackage{xspace}
\usepackage{xcolor}
\usepackage{amsmath}
\usepackage{pifont}
\usepackage{amssymb}
\usepackage{caption}

\graphicspath{{figs/}{figures/}{pictures/}{images/}{./}} 

\newtcolorbox{quotebox}[1][]{
  colback=gray!5,
  colframe=gray!60,
  fonttitle=\bfseries\small,
  title={#1},
  boxrule=0.5pt,
  arc=2pt,
  left=6pt,
  right=6pt,
  top=4pt,
  bottom=4pt,
  fontupper=\small\itshape
}

\newif\ifshowchanges\showchangestrue

\newcommand{\SysName}[0]{Living Dashboard\xspace}
\newcommand{\LD}[0]{Living Dashboard\xspace}
\newcommand{\Baseline}[0]{Baseline\xspace}

\newcommand{\greentick}{%
  \tikz[baseline=-0.5ex]{
    \node[
      fill=green!15,
      rounded corners=3pt,
      inner sep=3pt
    ] {\textcolor{green!50!black}{\checkmark}};
  }%
}
\newcommand{\redwrong}{%
  \tikz[baseline=-0.5ex]{
    \node[
      fill=red!15,
      rounded corners=3pt,
      inner sep=3pt
    ] {\textcolor{red!60!black}{$\times$}};
  }%
}

\newcommand{\greyskip}{%
  \tikz[baseline=-0.5ex]{
    \node[
      fill=gray!15,
      rounded corners=3pt,
      inner sep=3pt
    ] {\textcolor{gray!60!black}{$-$}};
  }%
}

\PassOptionsToPackage{hyphens}{url}

\begin{document}
\title{Living Dashboards: Automatically Self-Updating Visualization Dashboards}

\author{%
  Mingyu An\textsuperscript{1} \quad
  Heyon Jeon\textsuperscript{1} \quad
  Sungbok Shin\textsuperscript{2} \quad
  Jinwook Seo\textsuperscript{1} \\
  \bfseries
  Eduard Gr\"{o}ller\textsuperscript{3,4} \quad
  Niklas Elmqvist\textsuperscript{5} \quad
  Vaishali Dhanoa\textsuperscript{3,5}\thanks{Correspondence: \texttt{dhanoa@cs.au.dk}.}
  \\[4pt]
  {\small
    \textsuperscript{1}Seoul National University \quad
    \textsuperscript{2}Sogang University \quad
    \textsuperscript{3}TU Wien \quad
    \textsuperscript{4}VRVis GmbH \quad
    \textsuperscript{5}Aarhus University}
}

\maketitle

\begin{abstract}
    Visualization dashboards are widely used interactive tools, but a disconnect exists between the dynamic data they display and their static structure.
    End-users cannot modify the dashboard to answer new questions.
    We introduce \textit{Living Dashboards}, whose views are born, wither, revive, and die in response to how they are used.
    Rather than requiring manual reconfiguration, a living dashboard observes interaction and natural-language queries to autonomously wither neglected views and revive those used again.
    More consequential decisions, such as adding or retiring views, are deferred to the user.
    We formalize the concept as a four-dimensional design space and implement it in \SysName{}, a web-based prototype.
    We evaluate it in an exploratory between-subjects study ($N=12$) against an AI-supported baseline on analytical tasks.
    \LD{} participants answered more tasks correctly, reported lower workload, and rated the system higher on usability, though the two conditions differed in more than adaptive behavior alone.
\end{abstract}


\begin{center}
  \includegraphics[width=\textwidth]{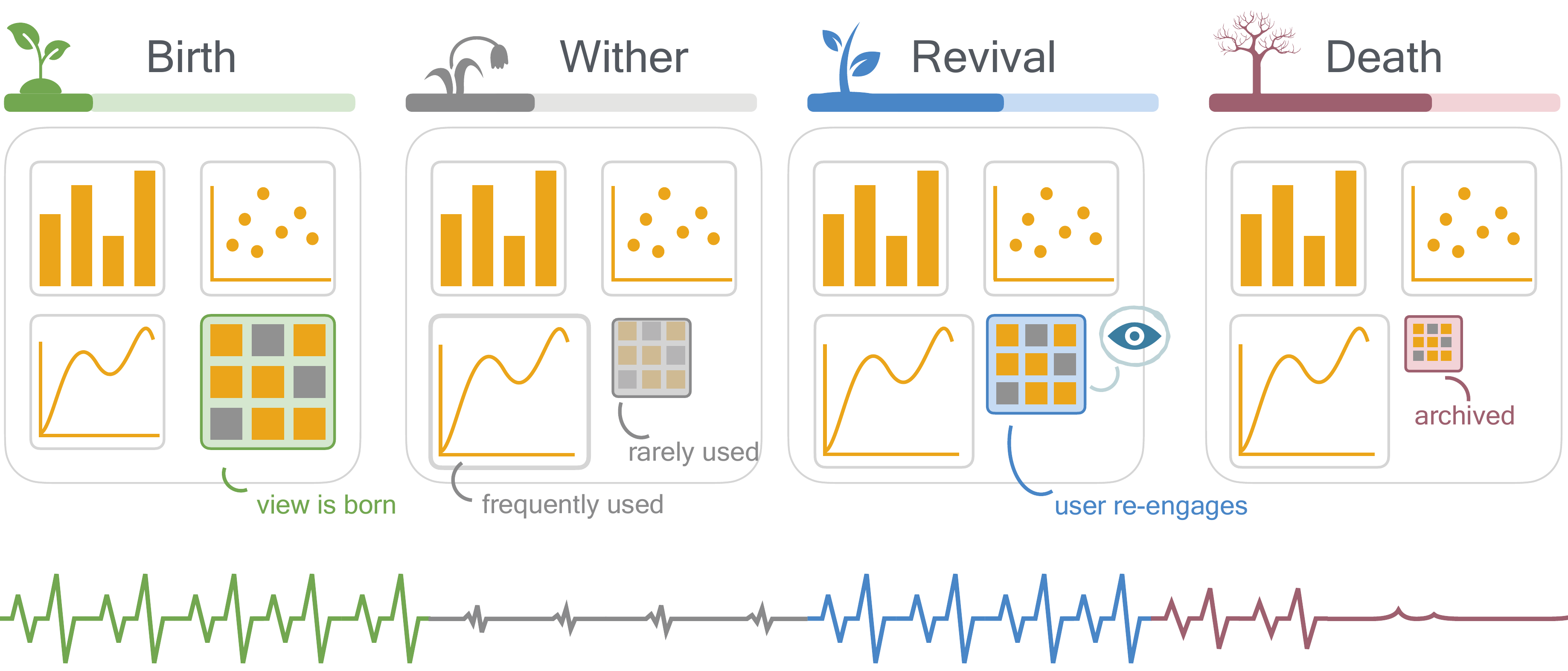}
  \captionof{figure}{%
    \textbf{Living Dashboard Lifecycle.}
    A living dashboard view passes through four stages driven by observed user attention.
    Birth: a new view is created at full prominence alongside existing views.
    Wither: neglected views shrink and fade (with a vignette effect) while the active view remains prominent (both due to user interaction and attention).
    Revival: a previously withered view regains size and visibility as interaction resumes.
    Death: a view that is constantly neglected is flagged for removal.
    The heartbeat trace below reflects the view's life status at the time.
  }
  \label{fig:teaser}
\end{center}

\section{Introduction}

Visualization dashboards are paradoxically both dynamic and static at the same time.
They are dynamic in their data: dashboards monitoring a chemical plant's reactor temperatures, an agency's ad revenue over the past 24 hours, or an e-commerce platform's daily sales can stream fresh values continuously~\cite{sarikaya_what_2019, few_information_2006}.
Yet many remain static in their structure: the layout, the set of views, and the visual encodings are fixed at design time and rarely change afterward.
This means that when a user's analytical questions evolve---as they often do during exploratory analysis, status meetings, or incident response---the dashboard may struggle to keep pace without manual changes, even as the numbers inside it continue to update.
Practitioner accounts describe how dashboards can decay into irrelevance as the questions they were built to answer become stale, prompting organizations to create new dashboards rather than maintain existing ones~\cite{oracle_dashboards_dead_2023, atlan_dashboard_sprawl_2026}.
Within an existing dashboard, this raises the question of how its individual views can remain useful as users' analytical needs change.

In this paper, we propose \textit{Living Dashboards}: visualization dashboards whose views are born, wither, revive, and die in response to users' needs and observed use.
Like living organisms, a living dashboard's components are not static: views are born at full prominence but wither in size and visibility when neglected, and revive to their original state when interaction resumes through clicks, filters, and selections.
The dashboard takes certain actions autonomously, such as withering or reviving views based on a focus score derived from these observed interactions.
It recommends more consequential changes, such as adding a new view or retiring one that has withered and remained unused, but applies them only with the user's approval.
In this way, it stays responsive to the people who use it rather than fossilizing around the intentions of the person who designed it.

However, interaction is only a proxy for attention: a view may receive little direct interaction and still be useful for comparison or occasional reference.
A living dashboard must therefore preserve the user's spatial mental map of the interface, ensure that changes are perceptible rather than silent, and keep every view---even demoted ones---accessible for reading and comparison.
We formalize these constraints alongside a four-dimensional design space---\textit{Observation}, \textit{Adaptation}, \textit{Temporality}, and \textit{Agency}---that characterizes the range of possible living dashboard behaviors.

We conducted a formative study with $N=4$~participants---three dashboard authors and one dashboard user---to understand current authoring and maintenance practices and inform the design goals that ground our design space.
We then implemented the concept in a web-based system called \textsc{\SysName} that dynamically adapts dashboard views as their use changes.
Finally, we evaluated the prototype in an exploratory between-subjects study with $N=12$ participants who performed analytical tasks with either \SysName{} or a baseline that offered AI guidance but no adaptive behavior and no AI-assisted view creation.
Participants using \SysName{} answered more tasks correctly and rated its usability more highly than those using the baseline, though the comparison reflects the combined effect of adaptation and AI assistance rather than adaptation alone.
Responses to the withering effect were mixed: participants found it helpful for focused, single-view analysis but frustrating when a task required comparing several views.

In summary, this paper contributes
(a) formative interviews with dashboard authors and a user that motivate our design goals;
(b) the \textit{Living Dashboards} concept, including a design space that characterizes adaptive dashboard behavior along four dimensions;
(c) a web-based implementation of the concept in the \textsc{\SysName} system; and
(d) an exploratory between-subjects study ($N=12$) of task performance, usability, and participants' responses to changing view prominence.

\section{Related Work}

The \LD{} lifecycle behavior draws inspiration from three areas of prior work: dashboard authoring and visualization recommendation, adaptive interfaces and mixed-initiative systems, and attention and focus in visualization.

\subsection{Dashboard Authoring and Recommendation}
Dashboards are often created by authors and deployed for users~\cite{dhanoa_d-tour_2025, sarikaya_what_2019}, which makes them effective for exploring the displayed data but can limit their usefulness when analytical questions evolve beyond what is displayed.
Authoring visualizations can be time-consuming and effort-intensive, but automated methods, and more recently LLMs, have reduced the effort required to create visualizations and dashboards.
VizML~\cite{hu_vizml_2019} uses machine learning models trained on large corpora of dataset-visualization pairs to predict design choices with accuracy comparable to human analysts. 
Draco~\cite{moritz_formalizing_2019} formalizes visualization design knowledge as logical constraints to enable automated chart assessment and generation. 
LADV~\cite{ma_ladv_2021} presents a deep learning approach that assists in authoring dashboard visualizations from images and sketches.
DMiner~\cite{lin_dminer_2024} mines design rules from existing dashboards to recommend the arrangement and coordination of views.
LightVA~\cite{zhao_lightva_2025} uses LLMs for visualization generation, among other tasks, with minimal user specification. 
Commercial dashboard platforms such as Tableau, Power BI, and Looker have also embraced AI-assisted authoring and usage by embedding features that generate visualizations from natural language queries. 

Such authoring support can help narrow the author-user gap discussed in prior work~\cite{walchshofer_transitioning_2024, dhanoa_d-tour_2025, dhanoa_process_2022}.
The tools above focus primarily on creating and recommending visualizations.
Our work extends this focus to what happens to views after they have been created: how they wither, revive, or are considered for retirement as their use changes.

\subsection{Adaptive Interfaces and Mixed-Initiative Interaction in Visualization Systems}

In the 1980s, Edmonds published several works in the area of adaptive systems, covering both user-triggered and self-adaptive interfaces that change automatically in response to experience with the user~\cite{edmonds_adaptive_1981, edmonds_mancomputer_1982}. 
To address the tension between user- and system-initiated adaptation, Horvitz~\cite{horvitz_principles_1999} formalized the principles of mixed-initiative interaction, proposing that the most effective systems elegantly couple automated services with direct manipulation, allowing both the user and the system to take initiative depending on context. 
Gotz and Wen~\cite{gotz_behavior-driven_2009} monitor a user's interaction patterns to infer their current analytical intent and recommend a better-suited view in response, demonstrating how observed behavior can guide visualization recommendation.
Findlater et al.~\cite{findlater_ephemeral_2009} introduce ephemeral adaptation, where predicted menu items appear immediately while the remaining items fade in gradually, reducing visual search time while preserving spatial consistency.
This use of gradual onset informs our own requirement that living dashboard transitions be perceptible without being jarring.
Gajos et al.~\cite{gajos_automatically_2010} present Supple, a system that automatically generates an entire personalized interface by modeling a person's devices, tasks, preferences, and abilities, showing improvements in speed, accuracy, and satisfaction for people with motor impairments compared with default interfaces.

Todi et al.~\cite{todi_adapting_2021} present a novel approach to adapting user interfaces that uses model-based reinforcement learning to plan sequences of adaptations, outperforming both a non-adaptive and a frequency-based policy. 
Cao et al.~\cite{cao_generative_2025} design generative and malleable user interfaces that evolve through a task-driven data model, demonstrating that interfaces can be continuously reshaped by generative AI in response to user activity.
In the space of visualizations, Yanez et al.~\cite{yanez_state_2025} survey the field of user-adaptive visualizations that can adapt to the characteristics and preferences of the user. 
They structure the process into five components: input capture, user modeling, user representation, adaptation assignment, and intervention. 
A recurring question their work brings up is whether the system or the user should drive adaptation. 
Lall\'{e} and Conati~\cite{lalle_role_2019} show that the effectiveness of customization in the visualization domain depends on user differences, including visualization literacy and locus of control, suggesting a role for system-driven assistance with customization.

Our work on Living Dashboards is inspired by prior work on adaptive interfaces and mixed-initiative systems, where visualizations adapt to user interaction, and both the user and the system take initiative to make the interface more malleable.
We bring these ideas to the lifecycle of dashboard views, combining automatic changes in prominence with user approval for additions and retirement.

\subsection{Attention and Focus in Visualization}

Attention is the cognitive process of selecting which sensory stimulus should be heeded, and accordingly has received significant attention over the years in human-computer interaction (HCI) and visualization research.
An early example was Bolt's seminal gaze-orchestrated windows~\cite{DBLP:conf/siggraph/Bolt81}.
The concepts of read wear and edit wear~\cite{Hill1992readwear} are forms of operationalized attention, making histories of reading and editing visible within an interface.
Since then, researchers have explored gaze for target acquisition~\cite{DBLP:conf/chi/StellmachD12}, multimodal interaction~\cite{DBLP:conf/ngca/StellmachSND11, DBLP:conf/uist/PfeufferACG14, DBLP:conf/sui/PfeufferMMG17}, and cross-device interaction~\cite{DBLP:conf/interact/TurnerABSG13}.

In visualization, there is a long history of using eye tracking to understand the user's attention for evaluative purposes~\cite{Burch_eyeTracking_11, Kim_eyeTracker_12, DBLP:conf/etra/BlascheckJKBE16, Kurzhals_16, Burch2021}.
More recent research explores gaze as an active input channel and the design of gaze-aware visualizations~\cite{srinivasan_attention-aware_2025, Jianu2025}.

Cockburn et al. examine how interfaces balance detail and context through overview+detail, zooming, focus+context, and visual cues that emphasize or suppress information~\cite{cockburn_review_2009}.
The withering and revival of dashboard views draw on this idea of changing visual prominence while retaining access to surrounding information.
In this work, we use mouse-based interactions as a proxy for users' attention and use these observations to adapt the behavior of the dashboard over time.
\section{Extracting Design Goals}

We conducted a formative study to understand current dashboard use and maintenance practices and set design goals to address the problems identified.

\subsection{Formative Study}

We conducted the study with four participants (A1--A4) to understand how dashboards evolve. A1 is a dashboard user, while A2, A3, and A4 are dashboard authors. Our goal was to examine how a dashboard is monitored after its launch and what happens to it once it no longer serves its original purpose.
All participants were recruited from the university. A1 works as a project controller; A2 and A3 work in the health department as a data consultant and project controller, respectively; and A4 is a consultant in the finance department.

Each interview was conducted online and lasted approximately 30 minutes. Before the study, participants signed a document covering GDPR and data privacy. We recorded and transcribed the interviews, extracted insights from the transcripts, and discussed these insights in a group meeting to inform the design goals.

\subsubsection{Monitoring Dashboard Usage}

A2, A3, and A4 regularly create dashboards for multiple users, and each monitors their evolution differently.
A2 relies on users to inform him when a dashboard is no longer in use, and he also checks periodically to determine whether disuse stems from a technical barrier or simply from the dashboard no longer being needed. A3 estimated that around 70\% of created dashboards are not in active use, while A4 put the figure at approximately 20\%. A4 has a formal system for monitoring dashboard usage, which he uses every two years to review and either delete or archive unused dashboards.
A1 noted that she actively uses only half of her dashboards, treating the other half as a reference library for occasional needs.

\subsubsection{Understanding Dashboard Decay}

Participants reported several reasons why dashboards fall out of use, including poor dashboard quality (A3), lack of onboarding or training (A3, A4), data quality issues (A2, A3), and users being unaware that a dashboard exists (A4).
A4 additionally pointed out that visualizations within a dashboard can fail to meet user needs or become overwhelming, which can also leave the dashboard unused.
Unused dashboards are handled in various ways: replaced by newer dashboards (A1), left in place if they cause no disruption (A2), retained due to lack of time for cleanup (A3), or archived for six months before deletion (A4). A4 also conducts ``live cleanup,'' actively refining visualizations within dashboards based on user feedback to ensure information is presented meaningfully.
Dashboard maintenance requires considerable time and effort. A3 redirects this investment toward building new dashboards, while A4 combines refinements based on user feedback with periodic reviews.

\subsubsection{Concerns Around Customized AI Dashboards}

Participants described dashboard development as a resource-intensive process (A2, A3, A4) and reported that users in their settings do not yet create dashboards on their own. For ad hoc analysis, users rely on Excel sheets (A1, A2, A3, A4) or SAP Business Objects (A2).
Participants raised several concerns about using generative AI tools such as Copilot to create dashboards. A1 cited data security and GDPR constraints as barriers and also worried that AI-generated dashboards could undermine standardization, leading each user to form their own view of the data and prolonging discussions in meetings. A2 doubted that Copilot-created dashboards would see regular use, questioned their quality, and noted that they could be difficult to develop further; since each new prompt may yield a different result, users would struggle to trust any single version. A3 and A4 believed users would not create their own dashboards regardless, citing a lack of time, skills, and data access.

\subsection{Design Goals}
Our formative study indicates that dashboard creation and maintenance require substantial effort, while ongoing maintenance is often limited. A4 pointed to their practice of refining visualizations within a dashboard, which indicates the need for keeping existing dashboards useful as users' needs change. A1 keeps the unused dashboards as a reference library, which also shows why infrequently used content may still be worth retaining. These accounts motivate support for dashboard reuse and the management of underused content (DG1, DG2).
Prior work on adaptive interfaces further informs how dashboards can respond to changes in use while preserving users' analytical context (DG3, DG4). Drawing on these findings and principles, together with our experience with dashboards, we propose the following design goals for the living dashboard concept.

\paragraph{DG1: Support Dashboard Reuse}

The system should help users address evolving analytical questions within an existing dashboard, prioritizing reuse of suitable existing views and supporting additions if the existing views do not support the user's needs.

\paragraph{DG2: Identify and Manage Underused Content}

The system should monitor view use and flag underused visualizations for review, giving users the autonomy to decide whether to retain or remove them.

\paragraph{DG3: Adapt Dashboards Through Use}

The system should respond to user interactions and progressively adapt the prominence of visualizations to accommodate shifts in exploration focus while preserving access to underused visualizations.

\paragraph{DG4: Preserve Dashboard Continuity}

The system should communicate how and why visualizations have evolved, allowing users to maintain their mental model of the dashboard and adapt to changes without losing analytical context.
\section{Design Space: Living Dashboards}
\label{sec:design-space}

The design goals motivate dashboards that can accommodate new analytical questions and changing patterns of use while remaining manageable over time (DG1, DG2, DG3).
The design challenge is to support this evolution while preserving users' orientation and analytical context (DG4).

\begin{quote}
    \textbf{Definition:} A \textit{living dashboard} is a visualization dashboard whose views are born, wither, revive, and die in response to users' needs and observed use.    
\end{quote}

Like a living document that evolves with its contributors, a living dashboard stays responsive to the people who use it rather than fossilizing around the intentions of the person who designed it.
The dashboard can act autonomously on presentation changes, such as withering or reviving views based on observed interactions, but defers to the user for consequential decisions such as adding or retiring a view.
Birth adds a view to the dashboard, while withering reduces its prominence and revival restores it as use resumes. Death refers to user-approved retirement from the active dashboard.
In this way, it draws on the idea of a \textit{read wear}~\cite{Hill1992readwear} user interface component.

\subsection{Design Constraints}
\label{sec:constraints}

Before describing the dimensions along which living dashboards can vary, we identify constraints that bound the space of acceptable designs.
These constraints are grounded in our design goals and in prior work on adaptive interfaces.

\paragraph{C1: Stable View Set}

DG1 suggests adding new visualizations if none of the existing visualizations answer the user's current analytical questions. Meanwhile, DG2 calls for flagging underused visualizations so that users can decide whether to retain or remove them. Additions and removals change the existing view set of the dashboard. The system must therefore ensure that these changes remain transparent and that the dashboard's analytical vocabulary is not altered without the user's approval. In other words, the view set remains stable until the user authorizes a change.

\paragraph{C2: View Prioritization with Continued Access}

To support adaptation through use (DG3), the system should respond to changing patterns of use---for example, by enlarging or foregrounding actively used views.
However, infrequently used views must remain accessible; adaptation should never render a view unreachable.
This echoes the classic adaptive interface trade-off: optimize the common case, but preserve the full capability set. 

\paragraph{C3: Mental Map}

Users build spatial memory of a dashboard's layout and rely on it for efficient navigation~\cite{DBLP:journals/fthci/ScarrCG13}.
To support DG4, adaptation should limit disruption to this spatial memory by preserving recognizable spatial references and avoiding unnecessary relocation of views. This does not mean the layout cannot change---it means changes must be gradual, predictable, or clearly communicated so that users can update their mental model without losing analytical context.

\paragraph{C4: Visible Change}

Adaptations that affect navigation or interpretation should be perceptible (DG3 and DG4), including changes to a view's location, size, encoding, or presence. Small, continuous adjustments may be conveyed through their overall transition or accumulated effect. Animated transitions, visual cues, or explicit notifications should help users recognize these changes. Where changes may occur outside the user's attention, cues should remain available afterward so that users can understand what has changed.

\paragraph{C5: View Legibility}

DG2 and DG3 allow adaptation to reduce a view's prominence---by shrinking it, moving it to the periphery, or muting its visual weight---but the content shown in each display state must remain legible.
A smaller view may show fewer details or use a compact representation, but its visible labels and marks should remain readable at the displayed size.
If the full visualization cannot be read at a reduced size, the system should provide a clear way to restore it for reading and comparison.
Blurring or obscuring a view prevents users from reading its content and should not persist when they engage with it.
The guiding principle: underused views should be \textit{quiet}, not \textit{illegible}.

\subsection{Design Dimensions}
\label{sec:dimensions}

We characterize the design space of living dashboards as a four-dimensional tuple: $\langle \textit{Observation}, \textit{Adaptation}, \textit{Temporality}, \textit{Agency} \rangle$.

Each dimension describes an aspect of a living dashboard's behavior, with choices in one dimension influencing the others.
We describe each in turn below.

\subsubsection[Observation]{\faEye~Observation}
\label{sec:observation}

Observation is the process by which a living dashboard collects evidence of use (DG1, DG2, DG3).
Three facets characterize what is observed.

\paragraph{Granularity}

Interaction events vary in specificity: from coarse signals such as opening a dashboard or scrolling, to fine-grained signals such as hovering over a specific mark, clicking a data point, or applying a filter.
Finer granularity can provide richer evidence of analytical intent but increases the complexity of the observation model.
In principle, observation could also include signals such as gaze or pointer proximity, as proposed in attention-aware visualization~\cite{srinivasan_attention-aware_2025}; we note this possibility but focus on mouse-based interaction in our prototype.

\paragraph{Scope}

Interaction with one view may imply engagement with others.
Cross-linked views that respond to a filter or selection in another view receive \textit{transitive} interaction: the user did not act on them directly, but their content changed as a consequence of the user's action.
A living dashboard must decide whether transitive interaction counts as use.
Counting it risks inflating the apparent relevance of linked views; ignoring it risks undervaluing views that serve a supporting role.

\paragraph{Attribution}

When multiple users share a dashboard, the system must decide whether to observe interaction individually or in aggregate.
Individual attribution enables personalized adaptation but may cause the dashboard to diverge across users.
Aggregate attribution enables shared adaptation---a form of social navigation~\cite{Dourish1994space}---where the dashboard reflects collective analytical focus.
Hybrid models are possible: aggregate by default, with individual overrides.

\subsubsection[Adaptation]{\faSlidersH~Adaptation}
\label{sec:adaptation}

Adaptation is the process by which a living dashboard modifies its content or presentation in response to users' needs and observed use (DG1, DG2, DG3).
We identify five forms of adaptation. Their intrusiveness depends on the extent of the change and the user's current task.

\paragraph{View Set}

The dashboard may add views that provide further analytical value or remove underused views. These operations change the content available within the dashboard and require explicit user approval (C1).

\paragraph{Emphasis}

The dashboard adjusts the relative size of views.
Heavily used views grow; underused ones shrink.
Size changes can preserve spatial arrangement, but must maintain access and legibility (C2, C3, C5).

\paragraph{Encoding}

The dashboard changes the visual representation of a view; for example, switching from a detailed table to a summary chart, or applying semantic zooming that reveals more detail as a view gains prominence.
This alters what the user sees within a view without moving it.

\paragraph{Layout}

The dashboard rearranges the spatial position of views; for example, promoting a frequently used view to a more prominent position.
This directly affects the user's spatial memory and must be handled with care to satisfy the mental map constraint (C3).

\paragraph{Visibility}

The dashboard controls how visually accessible a view is along a continuum of display states.
At full visibility, a view occupies its allocated space with all details rendered.
At reduced visibility, a view may be minimized to a compact representation---a labeled thumbnail, a sparkline summary, or a collapsed header---that signals the view's existence and content without occupying significant screen real estate.
At minimal visibility, a view may be reduced to a label or icon that can be expanded on demand.
In all cases, the view remains part of the dashboard (C1), can be restored on demand (C2), and has a representation that is legible at its displayed size (C5).
Although the view is never destroyed, reduced visibility may require user action to restore its details.

\subsubsection[Temporality]{\faClock~Temporality}
\label{sec:temporality}

Temporality governs when adaptation occurs and how observation ages (DG2, DG3).

\paragraph{Adaptation timing}

We distinguish when adaptations are applied from what triggers them.
\textit{Continuous} adaptation applies changes in real time as interaction accumulates, using animated transitions to maintain awareness.
\textit{Session-boundary} adaptation applies changes only between sessions, so the user encounters a modified dashboard upon return rather than during use.
\textit{Threshold-triggered} adaptation applies changes when a measure of use crosses a predefined threshold; thresholds can be checked continuously or at session boundaries.
Continuous adaptation responds during use but risks disrupting ongoing analysis. Session-boundary adaptation avoids changes during a session, but users still need to recognize and understand what has changed when they return (C3, C4).

\paragraph{Decay}

Without decay, observation accumulates so that early, heavily used views tend to dominate emphasis even after attention has moved on.
Decay reduces the influence of older activity, allowing the dashboard to track shifting analytical focus rather than ossifying around early usage patterns.
The rate of decay determines the dashboard's temporal memory: fast decay makes it reactive to recent activity; slow decay makes it reflect longer-term patterns of use.
Decay also determines what happens to views that were once prominent but are no longer used: they gradually recede, freeing emphasis for currently relevant views.

\subsubsection[Agency]{\faHandPaper~Agency}
\label{sec:agency}

Agency governs whether the user understands and can influence the dashboard's adaptation (DG2, DG3, DG4).

\paragraph{Control}

Control determines which adaptations the dashboard can apply autonomously and which require user approval.
Presentation changes can occur automatically in response to observed use, while additions and removals require explicit user approval (C1).
This allows the dashboard to respond to changing usage while keeping changes to its view set under user control.

\paragraph{Transparency}

A transparent living dashboard communicates what changed and why---for example, annotating a resized view with ``enlarged because you used this view 12 times this week.''
Transparency supports the mental map constraint by helping users update their expectations rather than merely noticing that something moved (DG4).
It also gives users a basis for judging whether an adaptation is appropriate for their needs.

\paragraph{Reversibility}

A reversible living dashboard allows users to override or undo adaptations---for example, pinning a view to prevent it from shrinking, or restoring a previous layout.
Reversibility shifts the system from autonomous adaptation toward mixed-initiative collaboration: the dashboard proposes changes, and the user accepts, rejects, or modifies them.
Full reversibility makes the system a tool; no reversibility makes it an agent.
The design goal of preserving analytical continuity (DG4) argues for at least partial reversibility.

\section{Implementation}
\label{sec:implementation}

\begin{figure*}[t]
  \centering
  \includegraphics[width=\linewidth]{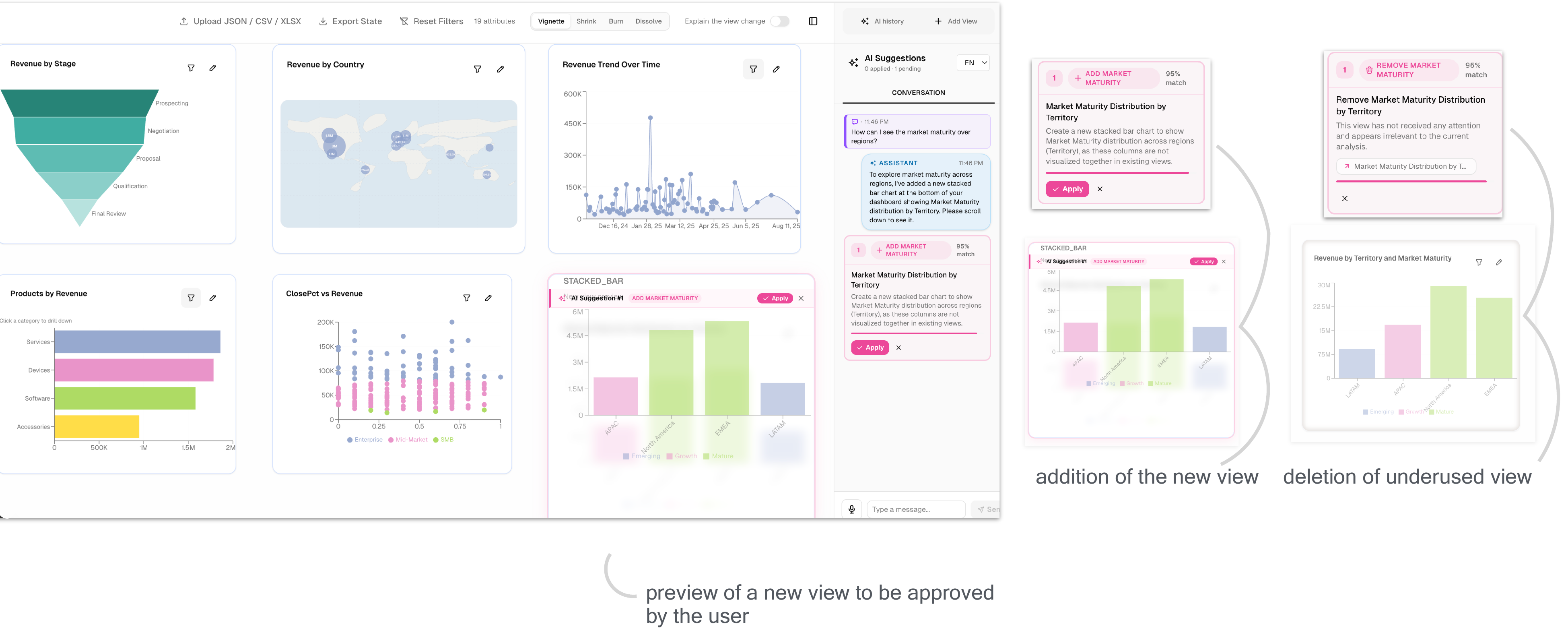}
  \caption{\textbf{The Living Dashboard Prototype.} The dashboard workspace is shown alongside the sidebar panel and header. A proposed view is previewed before the user approves its addition. An underused view can be recommended for removal, also subject to the user's approval.
  }
  \label{fig:system-figure}
\end{figure*}

We implemented \SysName{} as a web-based prototype (\url{https://living-dashboard.vercel.app/demo-dashboard}) that instantiates the design goals (DG1--DG4) and constraints (C1--C5) described in Section~\ref{sec:design-space}. The following subsections describe how the prototype: (i) reuses and adds views using the \textit{Sidebar Panel}, (ii) estimates use over time through a \textit{focus score}, (iii) withers and revives views based on that score, using our \textit{Layout Engine}, (iv) manages changes to the view set through a \textit{Recommendation Engine}, and (v) explains and reverses the changes it makes in the \textit{Dashboard Workspace}. Figure~\ref{fig:system-figure} shows the prototype and its recommendation interface.

\subsection{Reusing and Adding Views}

DG1 calls for supporting evolving analytical questions within an existing dashboard, prioritizing reuse of suitable views before introducing new content. To support analytical needs that cannot be inferred from interaction traces alone, the \LD{} prototype provides an AI chat interface in its Sidebar Panel through which users can pose questions in natural language. Its prompt includes the dataset schema and the current dashboard visualizations, together with a structured response format. We deliberately provide this information without the underlying data values, so that the AI serves as a guide for users conducting their own analysis, similar to the AI assistant created by Dhanoa et al.~\cite{dhanoa_hey_dashboard_2026}. This is an intentional design choice to encourage dashboard interaction and limit reliance on AI-generated numerical answers.
Based on the user query, the LLM first checks whether the analysis goals can be met by one or more existing dashboard visualizations.

When the LLM identifies suitable views, the system highlights them and increases their focus scores (described below). This connects query-based assistance to the dashboard's subsequent adaptation: the score reflects both observed interaction and boosts assigned in response to the user's query. If the LLM identifies a need for an additional view, the system recommends adding a visualization. Users can also configure a visualization directly through the \textit{Add New View} panel.

Because adding a visualization changes the dashboard's \textit{View Set}, the recommendation requires explicit user approval (C1). \textit{Birth} occurs when a user accepts an addition or creates a view directly. The recommendation process prioritizes reuse of existing views while preserving the user's ability to add new ones.

\subsection{Estimating Use Over Time}
DG2 and DG3 require the system to identify changes in view use and adapt as users' exploration shifts. We therefore instantiate the \textit{Observation} and \textit{Temporality} dimensions through a per-view focus score that combines recent interaction with the query-based boosts described above.

Within the \faEye~\textit{Observation} dimension, the prototype uses clicks, pointer movements, and dwell information to update view-level scores. Rather than treating all interactions uniformly, the system interprets interaction patterns at different levels of \textit{granularity}. 
For example, active mouse movements or clicks on a visualization are treated as signals of focused exploration. We combine their contributions as $u_i(t) = u_i^{\text{click}}(t) + u_i^{\text{pointer}}(t)$, where $i$ identifies the view and $t$ denotes time. Sustained low-speed movement is treated as a possible indication of passive attention, while transient pass-through movements are downweighted as low-intent interactions. A change in a visualization due to cross-filtering is not enough on its own to warrant a high focus score.
This is how our system defines the \textit{scope} of interaction with the visualization.

We weight the interaction input according to the view's current focus score:

\begin{equation}
\tilde{u}_i(t)=
\begin{cases}
u_i(t)\dfrac{f_0}{\max(1,f_i(t))}, & u_i(t)>0,\\
u_i(t), & u_i(t)\le0.
\end{cases}
\end{equation}
where $f_i(t)$ is the current focus score of view $i$, $f_0$ sets the scale of positive interaction updates, and $\tilde{u}_i(t)$ is the weighted interaction input.
This weighting gives lower-scoring views larger positive updates for the same interaction input, while higher-scoring views receive smaller increments. It supports the recovery of underused views when users re-engage with them and limits the additional influence of interactions with already highly focused views.

We then update the focus score over a time interval $\Delta t$:

\begin{equation}
\label{eq1}
f_i(t+\Delta t)
=
f_i(t)\cdot 0.5^{\Delta t/\tau}
+
\tilde{u}_i(t),
\end{equation}
where $\tau$ is the decay half-life, expressed in the same time units as $\Delta t$. The decay term reduces the influence of older activity, allowing the score to respond to changing patterns of use.

For \textit{attribution}, the prototype collects interaction from one active user at a time. A dashboard can be saved and loaded by a subsequent user, carrying its focus scores forward so that later interaction builds on the existing usage history.

\subsection{Withering \& Reviving Views}

DG3 calls for the dashboard to progressively adapt through use, while C2--C5 guide the design toward continued access, spatial continuity, perceptible changes, and legibility. Within the \faSlidersH~\textit{Adaptation} dimension, we therefore use \textit{Emphasis} and \textit{Visibility} for autonomous withering and revival, rather than automatically changing a view's encoding, spatial position, or presence.

We use a Layout Engine that maps focus scores to these two forms of adaptation. Lower-scoring views shrink and become less visually prominent, shifting visual emphasis toward views with higher scores. Renewed interaction increases their scores through the update mechanism described above, allowing them to regain size and visibility and thereby \textit{revive} (DG3). A revived view returns to its original size and full visibility; it does not grow beyond its initial allocation. Withered views keep their grid position, and the space a shrinking view vacates is left empty rather than redistributed among the remaining views, so that a view the user returns to reappears where they last saw it (C3). Visibility is linked to the focus score through:

\begin{equation}
\label{eq4}
\alpha_i = 0.25 + \frac{f_i}{1000} \cdot 0.75,
\end{equation}
where $\alpha_i \in [0.25, 1]$ is the visibility intensity and $f_i \in [0, 1000]$ is the focus score of view $i$.

The visual treatment of withering must balance two constraints: the change should remain perceptible so that users can recognize that a view has become underused (C4), while the visualization must remain legible and available for renewed interaction (C5). We therefore considered four visual encodings of decay (Figure~\ref{fig:underused-view}):

\begin{figure*}[t]
  \centering
  \includegraphics[width=0.8\linewidth]{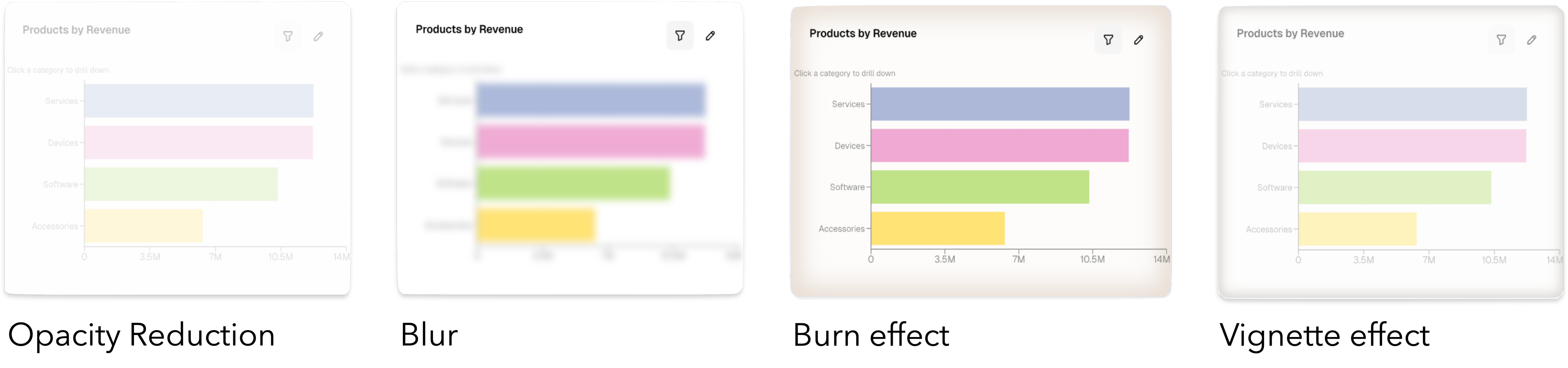}
  \caption{\textbf{Visual encoding options of view decay.} Opacity reduction, blur, burn effect, and vignette effect.
  }
  \label{fig:underused-view}
\end{figure*}

\begin{itemize}
    \item \textbf{Opacity reduction:} It decreases the visibility of a view by lowering its opacity. However, this can introduce a reinforcing feedback loop: less visible views may receive less interaction, leading to further decay.
    
    \item \textbf{Blur:} This applies a defocus effect to deemphasize a view. Similar to opacity reduction, blurred views become harder to interpret, potentially discouraging interaction and reinforcing underuse.

    \item \textbf{Burn effect:} To preserve visibility, we experimented with a burning paper effect, but we found that it made the view overly salient and potentially distracting.
    
    \item \textbf{Vignette effect:} We combined opacity reduction with the border treatment of the burn effect to implement a vignette effect that darkens the periphery of a view and subtly fades the visualization.
\end{itemize}
We chose the vignette effect, implemented as a subtle border shadow with reduced opacity, because we found it less distracting during development. To support readability, we restore full opacity as soon as the user hovers over the visualization. This restores contrast, while size recovery follows the focus-score updates described above. Together, these behaviors are intended to communicate withering while allowing users to return to the view for further exploration (C2, C5).

\subsection{Managing Changes to the View Set}
The \textit{Agency} dimension distinguishes adaptations that the system can perform autonomously from those that require user involvement. Adding or removing a view changes the dashboard's analytical content and therefore requires explicit user approval (C1).

For additions, we use a Recommendation Engine internally that proposes a visualization configuration, including its chart type, axis mappings, and title. For removal, it monitors focus scores and recommends removing a view when its score falls below the removal threshold. The recommendation brings underused content within the \textit{Dashboard Workspace} to the user's attention while preserving their decision over whether to retain or remove it (DG2).

Recommendations appear as cards containing the proposed change, its reason, and a preview. The system applies a recommended change to the view set only after approval. This implements the \textit{Control} facet of Agency for consequential dashboard changes. \textit{Death} occurs when the user approves the removal of a view from the dashboard.

\subsection{Explaining and Reversing Changes}
The prototype supports dashboard continuity (DG4) through explanations and recovery mechanisms. Each recommendation specifies both \textit{what} change should be made and \textit{how} the view should be updated. Its \texttt{reason} field explains \textit{why} the recommendation was generated, supporting the \textit{Transparency} facet of Agency and helping users understand changes to the dashboard (C4).

After a change is applied, the system maintains a history so that users can revert unwanted modifications, supporting \textit{Reversibility}. Together, these mechanisms allow the dashboard to adapt while keeping consequential changes understandable and recoverable (DG4).

This reversibility supports a mixed-initiative interaction paradigm: users remain in control while the system proactively suggests improvements.

\paragraph{Prototype Configuration.}
The AI chat uses GPT-4.1-mini~\cite{gptmini}, and the interface supports text and speech input in English and two additional languages. The header provides dataset and session controls. Focus-score parameters and thresholds are configured in the source code and reported in the supplementary material. The dataset used in the demonstration is described in the following section.
\section{Exploratory User Study}
\label{sec:evaluation}

We conducted an exploratory between-subjects study to understand how participants used the \SysName{} prototype for analytical tasks and responded to its changing views. We compared two conditions: the \textit{\LD{}} condition, in which participants used our prototype, and the \textit{\Baseline{}} condition, in which participants used a dashboard with AI guidance but without autonomous withering and revival. We included AI guidance in the baseline because it is available in commercial dashboard tools.
Both conditions used the same underlying AI model, but the \LD{} condition additionally supported AI-assisted view creation and highlighting of existing views (Table~\ref{tab:study-conditions}). The comparison therefore examines the experience of using the integrated prototype, including both its AI assistance and adaptive behavior.
For the \LD{} condition, we used an accelerated decay setting with a half-life of $\tau=45\,\mathrm{s}$ so that participants could observe withering during the sessions. The study focuses on short-term use and perceptions of the behaviors participants encountered.
We address three research questions:

\begin{itemize}
    \item \textbf{RQ1:} What differences in analytical task performance are observed between the Living Dashboard and Baseline conditions?
    \item \textbf{RQ2:} What differences in perceived usability and workload are observed between the two conditions?
    \item \textbf{RQ3:} How do users perceive and respond to the autonomous lifecycle behavior of the Living Dashboard?
\end{itemize}

\begin{table}[t]
    \centering
    \caption{Capabilities of the two study conditions.}
    \label{tab:study-conditions}
    \begin{tabular}{@{}lcc@{}}
        \toprule
        Capability & Living Dashboard & Baseline \\
        \midrule
        AI guidance & Yes & Yes \\
        AI-assisted view creation & Yes & No \\
        AI highlighting of existing views & Yes & No \\
        Autonomous withering and revival & Yes & No \\
        \bottomrule
    \end{tabular}
\end{table}

\begin{figure}[t]
    \centering
    \includegraphics[width=\linewidth]{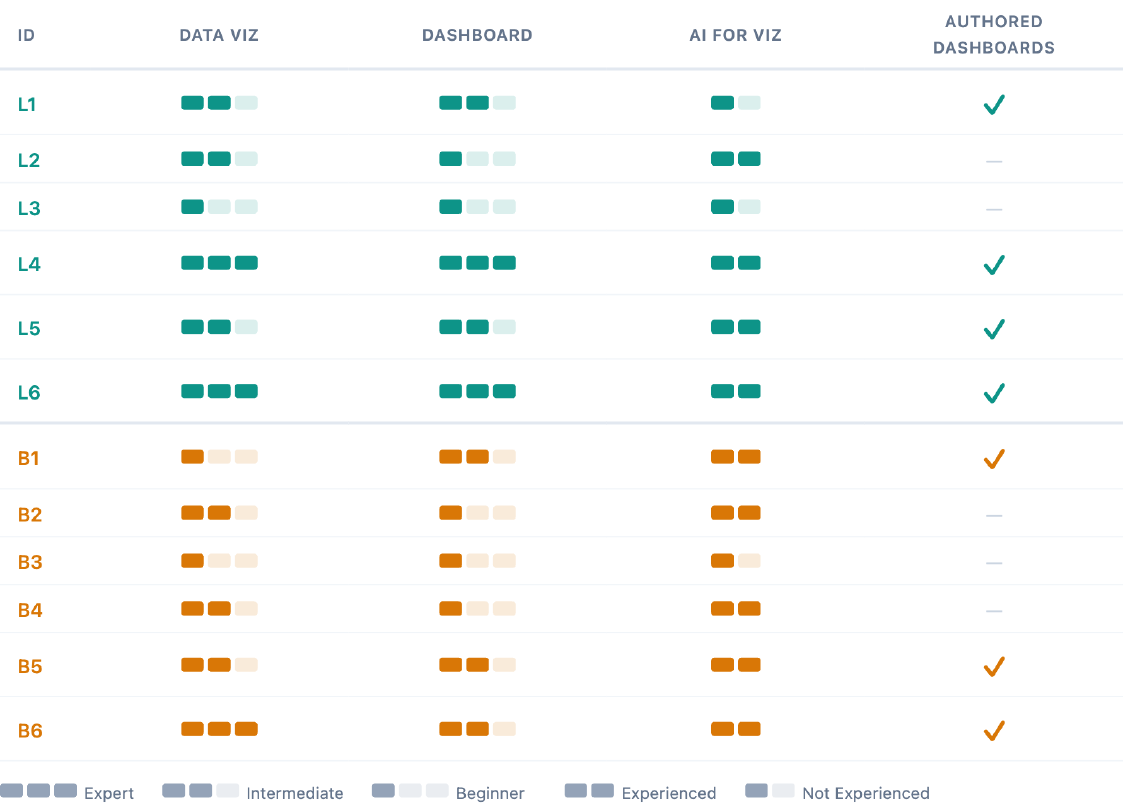}
    \caption{\textbf{Participants' Experience Levels.} Self-reported experience with data visualization (Data Viz), dashboard comprehension (Dashboard), chat interfaces for visualizations (AI For Viz), and authoring dashboards for other users (Authored Dashboards), collected before the study.}
    \label{fig:selfreport}
\end{figure}

\subsection{Participants}

We recruited $N = 12$ participants across two geographical regions, with six participants from each region. Participants were randomly assigned to conditions before their sessions: six to the Baseline condition (B1--B6) and six to the Living Dashboard condition (L1--L6), with three participants from each region per condition. We recruited them from university settings, including students from BSc to PhD levels and university employees, spanning experience levels from novice to expert in data visualization and dashboards. Before starting the tasks, participants self-reported their experience via a pre-questionnaire on a three-point scale (beginner, intermediate, expert) for both data visualization and dashboard usage. They also reported their experience with chat interfaces for visualizations (experienced, not so experienced) and whether they had created dashboards for users (yes, no) (Figure~\ref{fig:selfreport}). All participants provided informed consent and signed GDPR and data privacy documents before the study. We gave participants the option to complete the study in either English or a non-English language; six participants chose a non-English language.

\subsection{Dataset and Tasks}

We used a modified version of the Power BI revenue dataset~\cite{examplebi} with additional values to support a broader range of analytical tasks. Participants were presented with a preconfigured dashboard based on this dataset and were asked to complete ten analytical tasks. Eight of these ten tasks required exploration of existing views, while two (T4 and T9) required the creation of new views. The tasks were designed to reflect real-world analytical scenarios a dashboard user might encounter, ranging from straightforward information retrieval to more complex comparative analysis that required cross-filtering operations.
Participants received the following task list, numbered T1--T10 in Figure~\ref{fig:taskaccuracy}:
\begin{enumerate}
    \item[T1.] How much revenue is won by the Devices category?
    \item[T2.] List the revenue generated for each product category.
    \item[T3.] How many units were lost in the negotiation stage?
    \item[T4.] Which regions have the most mature markets?
    \item[T5.] Which campaign led to the most wins for the Manufacturing domain?
    \item[T6.] Which campaign was the first one to start in July 2025?
    \item[T7.] What was the revenue goal at the Proposal stage in the North America territory?
    \item[T8.] What are the subcategories in the Service category?
    \item[T9.] Which two industries contributed the most revenue won in Germany and Denmark?
    \item[T10.] In which time period did Australia's performance peak?
\end{enumerate}

\subsection{Procedure}
Each session was conducted individually and online, on the participant's own system, and lasted approximately one hour. After completing the pre-questionnaire, participants received a link to the prototype and were asked to select their assigned system, A (Living Dashboard) or B (Baseline), and share their screen. We introduced the dataset and demonstrated how to create views, interact with them, and use the AI features available in their condition. Participants then had a few minutes to familiarize themselves with the interface. In both conditions, we reminded participants to reset filters, reason through the tasks themselves, and phrase AI requests as questions rather than keywords. Participants in both conditions were allowed to skip tasks and complete them in any order, although all chose to follow the list from top to bottom.

Once participants felt comfortable with the interface, they worked through the ten analytical tasks using a think-aloud protocol. We recorded their screen, video, and audio. After the tasks, we interviewed participants about their experience and collected a post-questionnaire through LimeSurvey covering perceived workload, usability, and satisfaction. Living Dashboard participants were additionally asked to reflect on the concept itself. Participants saved their session and emailed the exported file to us for analysis; Living Dashboard participants used the ``Export State'' button for this step. Session state was not carried over between participants.

\subsection{Measures and Analysis}
We report task outcomes as correct, incorrect, or skipped. For each participant, we summarize the number of correct answers out of the ten assigned tasks; skipped tasks remain in this denominator and are shown separately from incorrect answers. We report participant-level summaries and task-level patterns descriptively.

The post-questionnaire included the ten-item System Usability Scale (SUS), TLX-based ratings of mental demand, effort, annoyance, and perceived success, and questions about overall satisfaction and the quality of AI suggestions. The four demand, effort, annoyance, and success ratings shown in Figure~\ref{fig:workfload} use a seven-point scale and are reported separately.
Living Dashboard participants also rated statements about autonomous behavior, visual clarity, and sense of control on a seven-point scale from strongly disagree to strongly agree, and compared their experience with a static dashboard on a scale from much worse to much better ($-2$ to $+2$).

The qualitative observations draw on participants' think-aloud accounts and the interviews conducted after the tasks. We transcribed the recordings, and two authors manually reviewed the transcripts and discussed their observations in a meeting. This review focused on how participants created views, responded to withering and revival, and encountered difficulties during analysis. We did not use a formal coding procedure.
Task completion time varied with participants' exploration styles, AI usage, and decisions to skip tasks. We do not use completion time to draw conclusions about efficiency in this study.

\subsection{Results}
We first describe participants' experiences of creating and exploring views, then report descriptive outcomes for task performance, questionnaire responses, and perceptions of adaptive behavior.

\subsubsection{Qualitative Results}

\paragraph{Creating New Views (Birth).}
The Living Dashboard condition supported the creation of new visualizations through the integrated AI feature, and participants engaged with it actively. We observed participants ask the AI to create particular charts---for instance, a bar chart filtered by a specific domain---rather than ask for the analytical answer outright. Participants expressed surprise when the AI generated a pre-configured view with the appropriate filters already applied. L6 remarked: \textit{"I don't have to think how I can combine the data to answer the question here"}. They also mentioned that this feature changed their approach to exploration: \textit{"it's so cheap to create a completely new chart specifically for my question"}. In the post-questionnaire, L1 noted that creating visualizations with AI support was helpful; L5 highlighted that AI-assisted visualization creation was a standout feature; and L6 specifically valued the speed at which AI could generate targeted visualizations.
In the Baseline condition, we observed participants turn to tables as a familiar option when creating views. B2 found table creation useful, while other participants stayed with chart types they already understood and encountered difficulties on tasks requiring custom views.

\paragraph{Exploring Existing Views (Wither and Revive).}
Withering and revival, which were present only in the Living Dashboard condition, received mixed reactions depending on the use context. Participants described benefits during tasks focused on a single view. L6 remarked during a task: \textit{"Oh it's nice, it is getting bigger when I look at it. It feels like it zooms in, makes it easier to find the right answer"}. L5 similarly found the reviving behavior useful when it worked in sync with the AI guidance, noting that the feature \textit{"was helpful when the AI guided my attention because it felt purposeful and intuitive"}. L6 described the withering effect as calming, making it easier to focus on the specific visualization. Interestingly, both L5 and L6 independently described the revival behavior as a zoom-in and zoom-out feature, suggesting that they interpreted the changes in size and opacity as a way of focusing on a view.
However, the same behavior became a source of frustration during comparison tasks that required simultaneously attending to multiple views. L5 described the experience as \textit{"cumbersome, especially when I needed to perform cross-dashboard comparisons"}, and at times characterized the decay effect as \textit{"an unnecessary animation"} when they needed to revisit a previously active view. L6 echoed this: \textit{"if I switch now my task and I ask another question, then of course it's a little bit harder to, let's say, dig up the respective chart again."}. We also observed that both L5 and L6 resorted to zooming into the web browser to read values on comparison charts, suggesting that a more explicit and faster revival mechanism could have supported these tasks better.

\paragraph{General System Observations.}
Across the two conditions, T6 (\textit{campaign starting in July 2025}) was among the tasks with the fewest correct answers.
We observed participants default to familiar visualization types, such as tables and line charts, while overlooking the Gantt chart available for this task. Even in the \LD{} condition, a suggestion to use a Gantt chart did not always lead participants to adopt it, as participants found it complex to explore.
We also saw participants struggle with cross-filtering and filtering in general, suggesting that more structured onboarding for these interaction techniques may be beneficial~\cite{dhanoa_process_2022, dhanoa_d-tour_2025}.

\subsubsection{Descriptive Quantitative Results}
\begin{figure}[t]
    \centering
    \includegraphics[width=\linewidth]{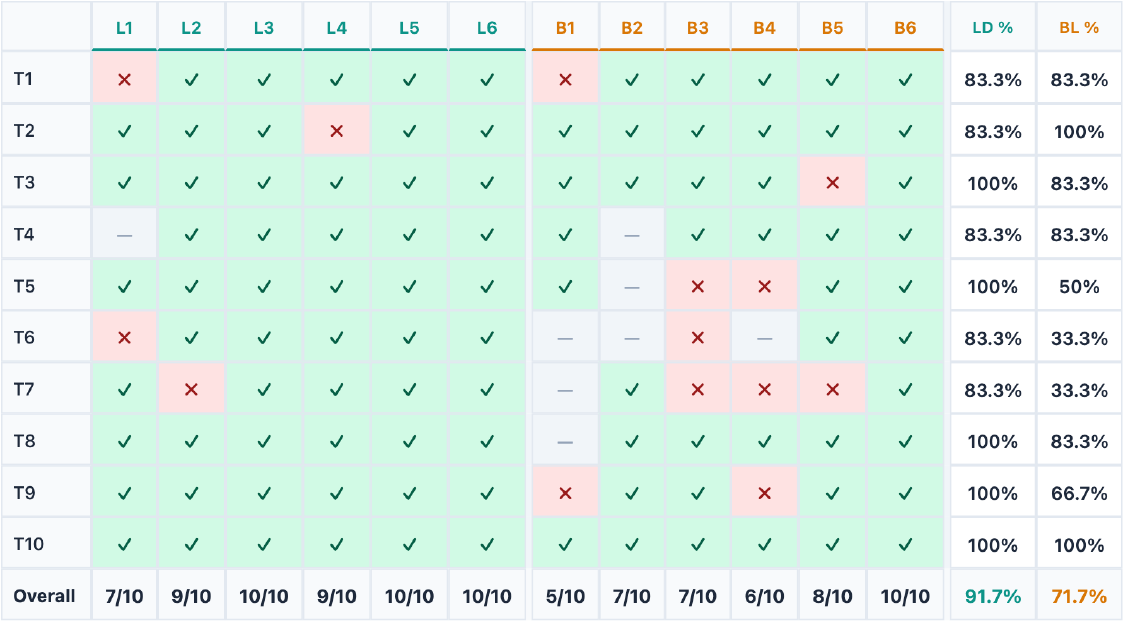}
    \caption{\textbf{Task outcomes by participant.} Each participant was assigned ten analytical tasks. Cells show outcomes: correct (green), incorrect (red), or skipped (grey). The last two columns show the percentage of correct answers for each task by condition; the bottom row shows each participant's total and the overall percentages.}
    \label{fig:taskaccuracy}
\end{figure}

\paragraph{RQ1: Task Accuracy.} Figure~\ref{fig:taskaccuracy} shows correct answers~\greentick, incorrect answers~\redwrong, and skipped tasks~\greyskip. Participants in the \LD{} condition answered an average of 9.17 tasks correctly (SD=1.17, range=7--10), compared to 7.17 (SD=1.72, range=5--10) in the \Baseline{} condition. Across the six participants in each condition, these outcomes correspond to 55 out of 60 assigned task instances answered correctly (91.7\%) in \LD{} and 43 out of 60 (71.7\%) in \Baseline{}.
The largest observed differences were in the exploratory tasks: T5 (\textit{campaign with most wins in Manufacturing}), T6 (\textit{campaign starting in July 2025}), and T7 (\textit{revenue goal at the Proposal stage in North America}). Out of the two view-creation tasks, T4 was tied between conditions and T9 favored \LD{}. 
Both conditions had the same number of correct answers on T1, T4, and T10 (\textit{Australia's performance peak}), while the baseline had one more correct answer on T2.

\begin{figure*}[t]
    \centering
    \includegraphics[width=\linewidth]{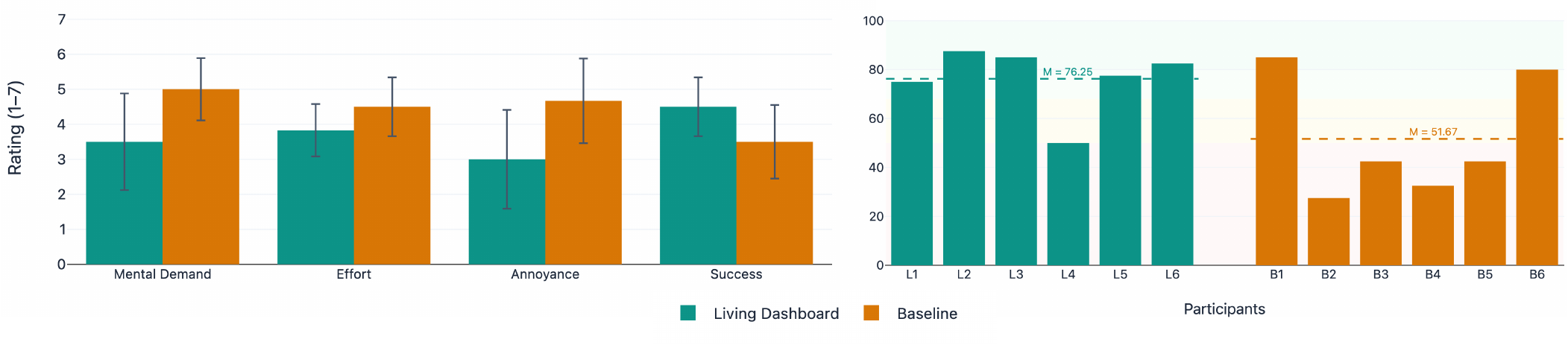}
    \caption{\textbf{Self-reported task experience (left) and System Usability Scale (SUS) scores (right).} Mental Demand, Effort, Annoyance, and Success were rated on seven-point scales. Lower values indicate less demand, effort, or annoyance; higher Success values indicate greater perceived success. Error bars show $\pm 1$ SD. The right panel shows individual SUS scores on a 0--100 scale, with dashed lines indicating condition means.}
    \label{fig:workfload}
\end{figure*}

\paragraph{RQ2: Perceived Demand, Effort, Annoyance, and Success.} Participants in the \LD{} condition gave lower mean ratings for mental demand (M=3.50, SD=1.38 vs. M=5.00, SD=0.89), effort (M=3.83, SD=0.75 vs. M=4.50, SD=0.84), and annoyance (M=3.00, SD=1.41 vs. M=4.67, SD=1.21) (Figure~\ref{fig:workfload}). Perceived success was higher in the \LD{} condition (M=4.50, SD=0.84) than in the \Baseline{} condition (M=3.50, SD=1.05). These ratings describe participants' experience of completing the tasks with each system.

\paragraph{RQ2: System Usability Scale (SUS).} The \LD{} condition yielded a mean SUS score of 76.25 (SD=13.67), compared to 51.67 (SD=24.63) in the \Baseline{} condition (Figure~\ref{fig:workfload}). Five out of six \LD{} participants scored 75 or higher, while the remaining participant scored 50.0. \Baseline{} scores were more variable, ranging from 27.5 to 85.0.

\paragraph{Satisfaction.} Mean overall satisfaction with the system was higher in the \LD{} condition (M=5.50, SD=0.55) than in the \Baseline{} condition (M=4.33, SD=1.03). Mean ratings of the perceived quality of AI suggestions also favored the \LD{} condition (M=4.33, SD=1.37 vs. M=2.83, SD=1.83). In \LD{}, these suggestions could also lead to the creation of new views or highlighting of existing ones.

\paragraph{RQ3: Living Dashboard Perception.} The six \LD{} participants gave their highest mean rating to the visual communication of active and inactive views (M=6.50, SD=0.55), followed by their reported understanding of layout changes (M=6.33, SD=0.52) and perceived sense of control during view changes (M=6.17, SD=0.75). The shrinking behavior received the most mixed responses regarding both its appropriateness for the workflow and its ability to help participants focus. Three of the six participants agreed that shrinking felt appropriate for their workflow (M=4.33, SD=1.86). Four agreed that shrinking helped them focus, while two disagreed (M=5.17, SD=2.14). These separate ratings reflect the mixed reactions described in the qualitative findings.
When asked to compare \LD{} with a static dashboard, five of the six participants rated it as better (mean=1.0), with L5 rating it as somewhat worse ($-1$). Because participants used only one condition, this item captures their perceived comparison with a static dashboard, rather than a preference based on using both study systems.

\subsection{Scope of the Findings}
The study provides an initial account of how participants used the prototype and responded to changes in view prominence during analytical tasks. The \LD{} condition combines AI-assisted view creation, highlighting of existing views, and autonomous withering and revival (Table~\ref{tab:study-conditions}), so the comparison does not identify which of these features individually accounts for the observed differences in task performance and ratings.
With six participants per condition, differing levels of expertise, a single dashboard, and brief sessions, the findings describe the experiences of this sample. 
No retirement recommendation was observed during the sessions, so the study does not assess the Death stage or longer-term effects on dashboard maintenance. The responses to withering and revival should also be understood in the context of the accelerated decay setting used in the study.
\section{Discussion}
In our exploratory study, participants using \SysName{} answered more tasks correctly and reported higher usability and lower workload than those using the \Baseline{}. Because the two conditions differed in both adaptive behavior and AI assistance, and because the study was small and short, we read these results as an early signal rather than a controlled effect of the living dashboard concept. The qualitative findings are more informative: they show how participants used AI and how they responded to views changing prominence, and they surface tensions that should inform future living dashboard designs. In this section, we reflect on these findings, discuss design alternatives and limitations, and outline directions for future work.
\subsection{AI as an Onboarding Assistant and Validator}
During the user studies, we observed that participants used AI as an \textit{onboarding assistant} and as a \textit{validator}, in addition to the analytical guide it was intended to be. Especially as users were getting accustomed to the dashboard, they used AI as an onboarding assistant, asking questions about what was displayed on each axis, the units of a specific column, and so on. Even when they had the answers, they used AI as a validator to confirm their own reasoning before committing to a response. L2, for example, used drill-down correctly and then asked the AI to validate their approach. L6 similarly used AI to double-check answers against the underlying data. L5 likened the experience to \textit{"riding an e-bike with pedal assist---it held my hand just right, instead of spoon-feeding me"}. These observations are from the Living Dashboard condition, and they align with findings by Dhanoa et al.~\cite{dhanoa_hey_dashboard_2026}, who use AI as a multimodal onboarding assistant.

\subsection{Design Alternatives for the Withering Effect}

The withering effect led to the most mixed responses in our study (in the \LD{} condition), especially during comparison tasks that required simultaneous access to multiple views as those views withered. Because a shrinking view keeps its position and the freed space is left empty (Section~\ref{sec:implementation}), withering created gaps on the dashboard rather than reflowing the remaining views. This was an intentional choice to preserve the user's spatial mental map, but it frustrated participants who wanted to rearrange views themselves. L5 noted: \textit{"I would have preferred more direct control. Features such as the ability to reorder tiles or manually zoom into a dashboard at full scale would have significantly improved the experience. I particularly felt the absence of these controls during comparison tasks, as I had to repeatedly scroll between tiles, which disrupted my workflow"}. L5 also described the dimming as \textit{"an unnecessary animation"} during comparison tasks, and both L5 and L6 resorted to zooming into their browser to read values on views that had withered.
 
We identify two ways to address these issues. First, \textit{layout reflowing}, either triggered by the user or by the system between sessions, could redistribute freed space among the remaining active views, keeping the layout dense and readable at all times. Second, more \textit{explicit revival controls}, for instance, through a double-click or a dedicated revival button, could give users a faster way to restore a withered view to full size without waiting for the focus score to rise naturally. The current prototype does not support manual reordering of views: it adapts the presentation of an author-defined layout rather than letting users restructure it. The \textit{Layout} dimension of our design space (Section~\ref{sec:design-space}) admits such changes, but we did not implement them.

As specified earlier, the withering effect in our study was deliberately accelerated to ensure participants could observe the lifecycle within the short duration of the sessions. In real-world deployments, the decay rate would be considerably slower, reflecting hours or days of inactivity rather than minutes. This means the frustration observed during comparison tasks may be less pronounced in practice, as users would have more time between interactions before views begin to wither noticeably.

\subsection{Interaction as a Signal for View Prominence}

A living dashboard adapts on what it can infer from use, so the quality of the \textit{Observation} signal bounds how well it can behave. Our prototype infers prominence from mouse interaction: clicks, pointer movement, dwell, and filters. Our exploratory study shows where this could be further improved. During comparison tasks, participants needed several views at once, but a view that is only being read produces little interaction, so the system withered views that were still in use. We noted this risk when introducing the concept; in practice it was the main source of friction.

Two design opportunities are possible. The system could observe more: dwell, pointer proximity, or gaze would capture attention that clicks and filters miss, at the cost of a more complex and less predictable model. Alternatively, it could accept that inferences will sometimes be wrong and make them cheap to correct, which is the motivation for the explicit revival controls discussed above.

\subsection{Use of Living Dashboard in Collaborative Settings}
\label{sec:collaborative}

The \SysName{} was designed for both independent analysis and collaborative settings. However, in our evaluation, we focused on individual use rather than a collaborative setting to avoid confounding factors such as group communication, coordination, and decision-making, which could have complicated the study design and the interpretation of the results.

We also developed a voice input feature that we did not explicitly introduce during onboarding, as it was primarily designed for collaborative scenarios. However, L6 independently discovered it and found it intuitive and easy to use. Our intention for the collaborative scenario was for the \LD{} to listen to the ongoing discussion in a meeting and autonomously generate or revive views based on the conversation's context. This could lead to fewer dashboard iterations and, perhaps, increased productivity during the meeting by enabling the on-demand creation of new views. We plan to evaluate this scenario in the future to investigate how different input signals from multiple participants can result in a coherent lifecycle behaviour.

\subsection{Limitations and Future Work}

Our work has several limitations across its design and evaluation that should be considered when interpreting the results. 

\paragraph{Number of Participants.} The sample size of $N=6$ per condition limits the generalizability of our quantitative findings. We would need a larger sample and a more controlled study design to draw stronger conclusions from the quantitative results. Therefore, in the future, we plan to conduct a large-scale evaluation of our work, incorporating the feedback from our current study results.

\paragraph{Generalizability of the Results.} The study was conducted on a single dataset and dashboard to ensure that no participant gained an advantage through domain familiarity. However, this dashboard does not represent the diversity of domains and data types in which dashboards are used in practice. We plan to integrate \SysName{} into existing dashboarding tools, such as Tableau and Microsoft Power BI, to evaluate the living dashboard concept across diverse real-world dashboards and datasets. 

\paragraph{Configuring Implementation Parameters.} Currently, our system requires authors to manually configure the implementation parameters for the decay rate, interaction events, and other events directly in the code. In the future, we plan to create a user interface that allows authors to configure these values more intuitively.

\paragraph{Decay Rate.} We accelerated the decay rate to ensure participants could observe the withering effect within the session duration. This likely amplified frustration during comparison tasks, as views withered much faster than they would in real-world deployments, where inactivity is measured over hours or days rather than minutes. A longitudinal study would be needed to investigate how users respond to slower and more natural decay rates over time.

\paragraph{Lack of Evaluation of the Death Stage.} Due to the short duration of the study, the death stage of the lifecycle was not observed — no views accumulated sufficient inactivity to trigger a retirement recommendation. Similar to the decay-rate observation, we would also need longitudinal studies to evaluate how users respond to retirement recommendations over extended periods of use and whether they understand the lifecycle behavior in its entirety.

\paragraph{AI Hallucination.} This was observed in both conditions and affected task performance in a few cases, especially for tasks T6 and T8, where the AI suggested out-of-order interactions that confused the participants. Some participants, especially self-reported experts, used existing views or KPIs to verify the information, while others relied on AI despite repeated failures. Since both conditions used the same underlying AI, hallucination represents a shared limitation of the current implementation. This could be addressed in the future by introducing dataset- or visualization-level constraints to reduce hallucinations in dashboard-specific analytical contexts.
\section{Conclusion}

In this paper, we introduced \textit{Living Dashboards}, a concept for visualization dashboards whose views are born, wither, revive, and die in response to how they are used.
Unlike static dashboards, which preserve the structure their designer intended regardless of whether it still serves their users, a living dashboard evolves by responding to use rather than design alone.
We formalized this concept through a four-dimensional design space---Observation, Adaptation, Temporality, and Agency. It has been implemented in \SysName{}, a web-based prototype that manages the lifecycle of dashboard views through a focus score derived from observed interactions. 
The system acts autonomously on low-stakes changes, such as withering and reviving views, while deferring to the user for consequential decisions, such as recommending the addition or retirement of a view.
In an exploratory study with $N=12$ participants, those using \SysName{} answered more tasks correctly and reported lower workload and higher usability than those using the \Baseline{}, though the conditions differed in more than the adaptive behavior alone.
Participants understood and generally welcomed the view changes they experienced, withering and revival, and responded positively to the visual communication of view state and to AI-assisted view creation.
Withering drew the most mixed responses: helpful for focused single-view work, frustrating during comparison.

\section*{Use of Generative AI}

The authors used generative AI tools (Codex by OpenAI) to support coding for a substantial portion of the implementation work, and writing the python script for generating the visualizations in Figure~\ref{fig:workfload}. All outputs were reviewed and verified by the authors.

\section*{Acknowledgments}
    This work was partially supported by Villum Investigator grant VL-54492 by Villum Fonden.
    Any opinions, findings, and conclusions expressed in this material are those of the authors and do not necessarily reflect the views of the funding agency.

\bibliographystyle{ACM-Reference-Format}
\bibliography{living-dashboard}
\end{document}